\documentclass[3p,twocolumn]{elsarticle}
\usepackage{graphicx}
\usepackage{xcolor}
\usepackage{amsmath}
\usepackage{lineno}
\usepackage{url}
\journal{Journal Name}

\begin{document}

\begin{frontmatter}

\title{Ground-State Structure Search of Defective High-Entropy Alloys Using Machine-Learning Potentials and Monte Carlo Sampling}

\author[tamu]{Siya Zhu}
\author[tamu]{Raymundo Arr\'{o}yave\corref{cor1}}
\cortext[cor1]{Corresponding author}
\ead{rarroyave@tamu.edu}

\address[tamu]{Department of Materials Science and Engineering, Texas A\&M University, College Station, TX 77843}
\address[tamu_meen]{J. Mike Walker '66 Department of Mechanical Engineering, Texas A\&M University, College Station, TX 77843}
\address[tamu_msen]{Wm Michael Barnes '64 Department of Industrial and Systems Engineering, Texas A\&M University, College Station, TX 77843}

\begin{abstract}
Resolving the atomic-scale structure of defective high-entropy alloys (HEAs) containing interstitial species remains a major computational challenge due to the vast configurational space and the limitations of existing methods. Here we introduce PAIPAI (Package for Alloy Interstitial Predictions using Artificial Intelligence), a Monte Carlo framework coupled with machine-learning interatomic potentials (MLIPs) that searches for ground-state atomic configurations in HEAs with defects and interstitials. PAIPAI employs a dual-worker architecture---fast workers for rapid configurational screening and slow workers for high-accuracy refinement---coordinated through a shared waiting pool, enabling efficient parallel sampling. We demonstrate PAIPAI through three case studies: (i) surface segregation in a Ti--V--Cr--Re slab; (ii) interstitial oxygen and boron aggregation in bulk BCC Nb--Ti--Ta--Hf; and (iii) coupled metallic and interstitial segregation at grain boundaries in Nb--Ti--Ta--Hf. In all cases, Monte Carlo--optimized structures are significantly lower in energy than any configuration obtained by random sampling, and MLIP energy rankings are validated against density functional theory calculations. PAIPAI provides a general and efficient framework for predicting atomic ordering, segregation, and interstitial behavior in complex, defective HEA systems.
\end{abstract}

\begin{keyword}
high-entropy alloys \sep machine-learning interatomic potentials \sep Monte Carlo sampling \sep interstitial defects \sep grain boundary segregation
\end{keyword}

\end{frontmatter}

\section{Introduction}
High-entropy alloys (HEAs) have attracted increasing attention due to their exceptional mechanical, thermal, and chemical properties~\cite{yeh2004nanostructured, cantor2004microstructural, miracle2017critical, senkov2015accelerated}. Understanding their atomic-scale structures is fundamental to the exploration and optimization of HEAs. Various types of defects---including vacancies, dislocations, surfaces, and grain boundaries---play critical roles in determining mechanical behavior~\cite{li2021mechanical}. In addition, interstitial elements such as oxygen, boron, or carbon can significantly alter the stability and mechanical properties of HEAs~\cite{lei2018enhanced, zhou2024formation, wu2021short, wang2016effect, ko2025boron}. Experimentally, conventional techniques such as X-ray diffraction (XRD) and scanning electron microscopy (SEM) are widely used to characterize crystal structures and elemental distributions. However, these methods often provide only averaged or indirect structural information. Although some advanced methods, such as scanning transmission electron microscopy (STEM) and extended X-ray absorption fine structure (EXAFS), can provide atomic-scale configuration and short-range ordering information, they remain subject to their intrinsic methodological limitations, restricted statistical representativeness due to localized sampling, and substantial experimental cost~\cite{liu2025computational, he2024quantifying, zhang2017local, zhang2020short}. Consequently, theoretical and computational simulations are essential for resolving the atomic structures of HEAs.

Several computational approaches have been developed to construct and investigate the atomic structures of HEAs~\cite{liu2025computational, toda2017simulation}. Molecular dynamics (MD) simulations, for example, can directly model atomic motions and structural evolution~\cite{granberg2016mechanism}, but the accessible time scales are typically much shorter than those relevant to real physical processes, which may prevent accurate sampling of thermodynamically stable configurations. Special quasirandom structures (SQS)~\cite{zunger1990special} are commonly used to model HEA solid solutions by constructing structures whose correlation functions approximate those of a random solid solution, and have been widely applied in the calculation of solid-solution energies and phase diagram construction~\cite{van2017software, zhu2025accelerating, zhu2025machine}. However, real HEAs are rarely perfectly random: short-range order is often present~\cite{sheriff2024quantifying,cao2025capturing}, and elemental segregation can occur in the vicinity of defects such as surfaces, dislocations, and grain boundaries to reduce the total energy~\cite{ma2024mechanism, li2020grain}. These ordering and segregation effects are not captured by SQS models, which, by construction, enforce random-like correlations.

For investigations focused on total energies and thermodynamic ground states at the atomic level, density functional theory (DFT)~\cite{kresse1996efficient} is often employed. DFT provides high accuracy and detailed information on local distortions through structural relaxations, but its computational cost becomes prohibitive for HEAs, which generally require large supercells~\cite{odetola2024exploring}. Some studies rely on single-point DFT calculations without full relaxation to reduce the cost; however, the resulting unrelaxed structures may have significantly higher energies and fail to represent the true ground states~\cite{chang2024high,huang2024unraveling,luo2025determinants}. The cluster expansion (CE) method~\cite{sanchez1984generalized} offers an efficient alternative by fitting configurational energies to a model trained on DFT data, enabling rapid energy evaluations~\cite{van2002automating, nataraj2021systematic, zhu2023probing}. Nevertheless, the accuracy of CE depends on the quality of the training set and is typically limited to ideal lattice configurations, making it difficult to apply to systems containing defects~\cite{van2009multicomponent}. Overall, no existing method can simultaneously treat structural defects, accommodate interstitial species, and perform full structural relaxation across large configurational spaces at tractable computational cost.

Recently, universal machine-learning interatomic potentials (MLIPs)~\cite{focassio2024performance}, pretrained on large DFT databases, have emerged as powerful tools for atomistic simulations~\cite{rosenbrock2021machine, hodapp2021machine}. Unlike traditional empirical potentials such as the embedded atom method (EAM)~\cite{daw1993embedded}, which rely on fixed functional forms with limited transferability across compositions and chemical environments, MLIPs learn flexible representations of the potential energy surface directly from first-principles data, achieving near-DFT accuracy at a fraction of the computational cost. Foundation models such as GRACE~\cite{bochkarev2024graph}, MACE~\cite{batatia2022mace}, and CHGNet~\cite{deng2023chgnet} have demonstrated broad transferability across the periodic table, enabling simulations of chemically complex systems without system-specific fitting. For HEAs in particular, MLIPs have been successfully applied to solid-solution energy calculations and structural modeling within CALPHAD workflows, demonstrating their potential for large-scale exploration of multi-component alloy systems~\cite{zhu2025accelerating, zhu2025machine, kunselman2025construction}. However, MLIPs alone do not address the combinatorial challenge of exploring the vast configurational space of multi-component alloys containing defects and interstitial species; an efficient sampling strategy is still required.

In this work, we introduce the \textbf{Package for Alloy Interstitial Predictions using Artificial Intelligence (PAIPAI)}. The framework combines MLIPs with a Monte Carlo scheme to search for thermodynamically stable atomic configurations by minimizing the total energy at 0 K, as defined by the underlying MLIP or DFT reference. Given a predefined lattice containing both substitutional atomic sites and interstitial sites, PAIPAI evaluates the energy of each configuration after full structural relaxation and explores the configurational space using an atom-exchange Monte Carlo scheme to identify ground-state (i.e., lowest-energy) structures. Compared to MD, PAIPAI directly targets thermodynamic ground states rather than being limited by short dynamical time scales. Compared to SQS, it naturally captures clustering effects and short-range order. Compared to DFT, it enables fully relaxed simulations of hundreds of atoms with orders-of-magnitude higher efficiency, allowing the sampling of over 100,000 configurations. Compared to CE, PAIPAI can explicitly treat structural defects and interstitial species.

Furthermore, PAIPAI introduces a novel dual-worker architecture comprising fast and slow workers, coordinated via a waiting pool. Fast workers perform rapid, low-accuracy relaxations to efficiently screen configurations, whereas slow workers perform high-accuracy relaxations on selected candidates. Through dynamic communication with the waiting pool, this strategy enables highly efficient parallel execution and relaxes the strict sequential constraints of traditional Markov-chain Monte Carlo approaches, resulting in substantially improved sampling efficiency.

The remainder of this paper is organized as follows. In \textbf{Results}, we present three case studies of increasing complexity: surface segregation in a Ti--V--Cr--Re HEA slab without interstitials, interstitial aggregation and solubility estimation in bulk BCC Nb--Ti--Ta--Hf, and coupled metallic--interstitial segregation at a $\Sigma$5(120) grain boundary in Nb--Ti--Ta--Hf. We compare Monte Carlo--optimized structures with random sampling to quantify the approach's effectiveness. In \textbf{Discussion}, we summarize our conclusions and discuss future extensions of the framework. Finally, in \textbf{Methods}, we introduce the PAIPAI framework, including the Monte Carlo sampling scheme, the dual-worker architecture, and the MLIP and DFT calculation settings.

\section{Results}
\subsection{Surface segregation in a Ti--V--Cr--Re alloy slab without interstitials}
\label{sec:slab}
To illustrate the application of PAIPAI, we first investigate surface segregation in a HEA slab without interstitial species. Surface segregation---the preferential accumulation of certain elements at free surfaces---is a common phenomenon in multi-component alloys, driven by differences in elemental surface energies, atomic sizes, and chemical interactions. In HEAs, segregation can significantly affect surface-sensitive properties such as oxidation resistance, catalytic activity, and corrosion behavior. Several computational studies have addressed surface segregation in HEAs using a range of methods, including DFT calculations of surface formation energies~\cite{ferrari2020surface}, MC--MD simulations with empirical potentials~\cite{dahale2022surface, chatain2021surface}, and, more recently, alchemical machine-learning potentials coupled with Monte Carlo sampling~\cite{mazitov2024surface}. Here, we apply a similar MLIP--MC strategy within the PAIPAI framework to investigate surface segregation in a Ti--V--Cr--Re refractory HEA slab, demonstrating the ability of PAIPAI to identify nontrivial chemical ordering at surfaces through efficient configurational sampling.

The BCC slab model consists of eight atomic layers, containing a total of 128 atoms with a composition of Ti$_{19}$V$_{77}$Cr$_{26}$Re$_6$. Periodic boundary conditions are applied in the in-plane directions, while a vacuum region of 15 $\AA$ is introduced. The initial metallic configuration is constructed as a random solid solution. Monte Carlo sampling is performed over the chemical degrees of freedom of the metallic lattice. At each trial step, the chemical identities of two randomly selected metal atoms are exchanged. The resulting structure is subsequently relaxed using the MLIP. The relaxed total energy is evaluated, and the trial configuration is accepted or rejected according to a Metropolis criterion.

\begin{figure}
    \centering
    \includegraphics[width=\linewidth]{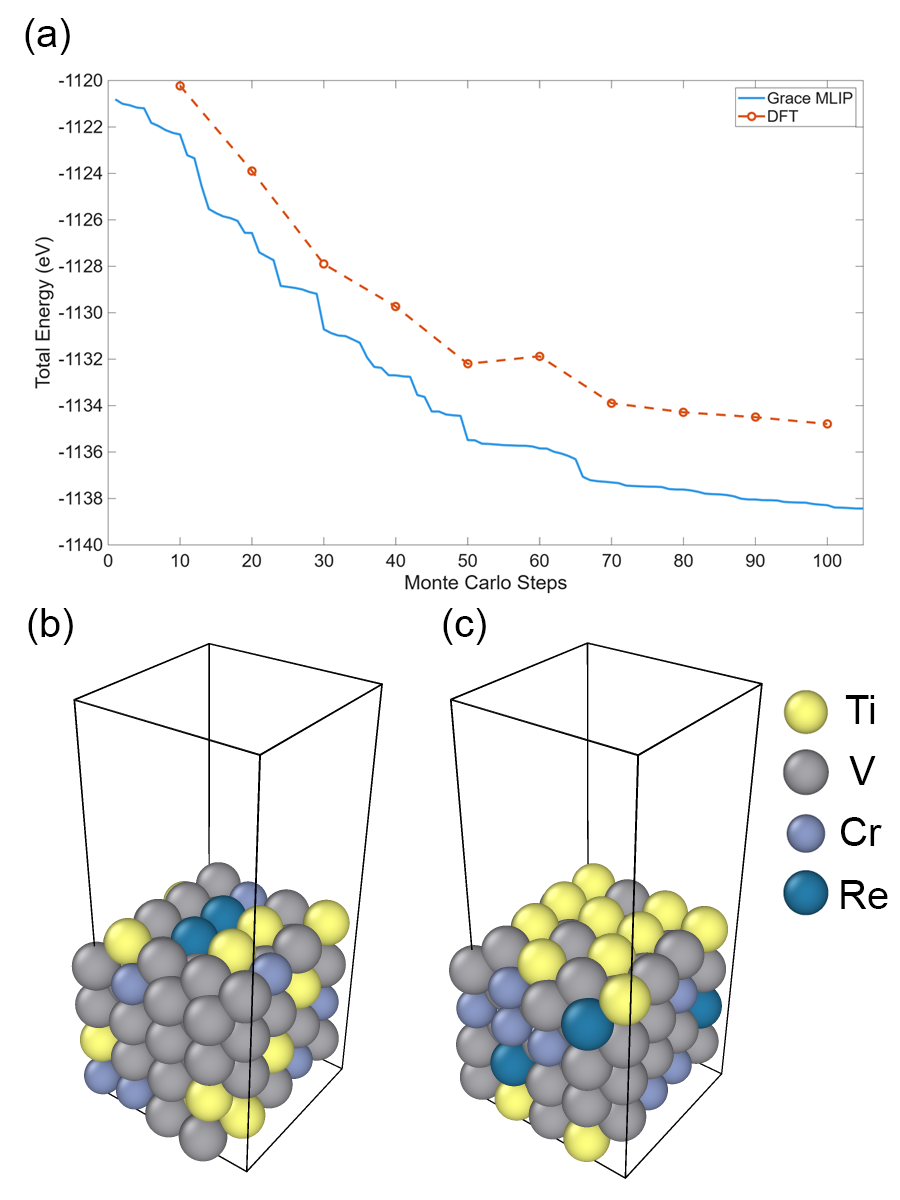}
    \caption{Monte Carlo process of Ti--V--Cr--Re slab. (a) Total energy of the slab versus Monte Carlo steps, with MLIP and DFT calculations; (b) initial random atomic structure; (c) atomic structure after $10^5$ accepted Monte Carlo steps with PAIPAI.}
    \label{fig:slab-mc}
\end{figure}

In Fig.~\ref{fig:slab-mc}(a), we illustrate the total energy of the slab versus Monte Carlo steps. Starting from a random configuration, the total energy decreases rapidly during the initial stage of the Monte Carlo process, with an overall energy reduction of approximately 20 eV. The energy subsequently converges after about $10^5$ accepted Monte Carlo steps, indicating that the system has almost reached a thermodynamically stable configuration under the present sampling conditions.

To assess whether the MLIP energies used in the Monte Carlo sampling reliably reflect real thermodynamic stability, we perform a direct comparison with DFT calculations. Configurations obtained from our MLIP-driven Monte Carlo simulation are extracted every 10 accepted Monte Carlo steps. Each selected structure is subsequently fully relaxed using VASP, and its total energy is evaluated. Both MLIP and DFT energies are illustrated in Fig.~\ref{fig:slab-mc}(a). While small quantitative differences between the MLIP and DFT energies are observed, these discrepancies originate from systematic errors inherent to the MLIP model, and the relative energy ordering of configurations is preserved. Both MLIP and DFT energies exhibit consistent convergence behavior along the Monte Carlo trajectory. Therefore, the identified thermodynamically stable configurations should remain unchanged when assessed at the DFT level.

In Fig.~\ref{fig:slab-mc}(b) and~\ref{fig:slab-mc}(c), we compare the initial random metallic structure with the configuration obtained after $10^5$ accepted Monte Carlo steps. A clear chemical segregation pattern emerges: titanium atoms show the strongest tendency to segregate to the top and bottom free surfaces, followed by vanadium, whereas chromium and rhenium preferentially remain in the slab interior. 

To quantitatively explain and support the surface segregation behavior, we compute the BCC (100) surface energies of Cr and V using DFT with the same slab geometry (pure BCC Ti and Re are unstable at low temperatures). For each element, the metallic species in the slab structure are replaced by a single pure element, and the corresponding bulk BCC energy is also evaluated. The surface energy of each element is obtained using:
\begin{equation}
    \gamma = \frac{E_\text{slab}-NE_\text{bulk}}{2A},
\end{equation}
where $E_\text{slab}$ is the total energy of the relaxed slab structure, $E_\text{bulk}$ is the energy of the BCC primitive cell, $N$ is the number of atoms in the slab structure ($N=128$ in our case), and $A$ is the lateral area of the slab simulation cell. The calculation of surface energies is summarized in Table~\ref{tab:surface_energy}.

\begin{table*}[htbp]
\centering
\caption{DFT calculated surface energies of the BCC (100) surface of V and Cr.}
\label{tab:surface_energy}
\begin{tabular}{lcccc}
\hline
 & $E_{\mathrm{slab}}$ (eV) & $E_{\mathrm{bulk}}$ (eV/atom) & $A$ ($\text{\AA}^2$) & $\gamma$ (J/m$^2$) \\
\hline
V  & -1102.8732 & -8.9458 & 138.9098 & 2.4328 \\
Cr & -1162.0845 & -9.4941 & 128.6040 &  3.3114\\
\hline
\end{tabular}
\end{table*}

The calculated surface energies are in good agreement with prior studies\cite{vitos1998surface, ossowski2008density} and indicate that V has a lower surface energy than Cr, suggesting that Cr atoms tend to avoid the free surfaces and preferentially remain in the slab interior. Although Ti and Re lack BCC surface energy data due to their instability, previous studies \cite{lee2018surface} have reported a parabolic dependence of surface energy across transition metals with d-electron filling due to the progressive occupation of bonding states up to half-filling and subsequent population of antibonding states, as described by the Friedel model of d-band cohesion\cite{friedel1976physics}.  This trend is consistent with the observed surface occupation behavior: Ti strongly enriches at the free surfaces, while V does not strongly avoid surface sites and can occupy the remaining surface positions due to the limited Ti concentration; in contrast, Cr and Re, with substantially higher surface energies, tend to remain away from the free surfaces.
The observed segregation pattern demonstrates that PAIPAI can recover physically meaningful chemical ordering from a fully random initial configuration through efficient Monte Carlo sampling.

\begin{figure}
    \centering
    \includegraphics[width=0.9\linewidth]{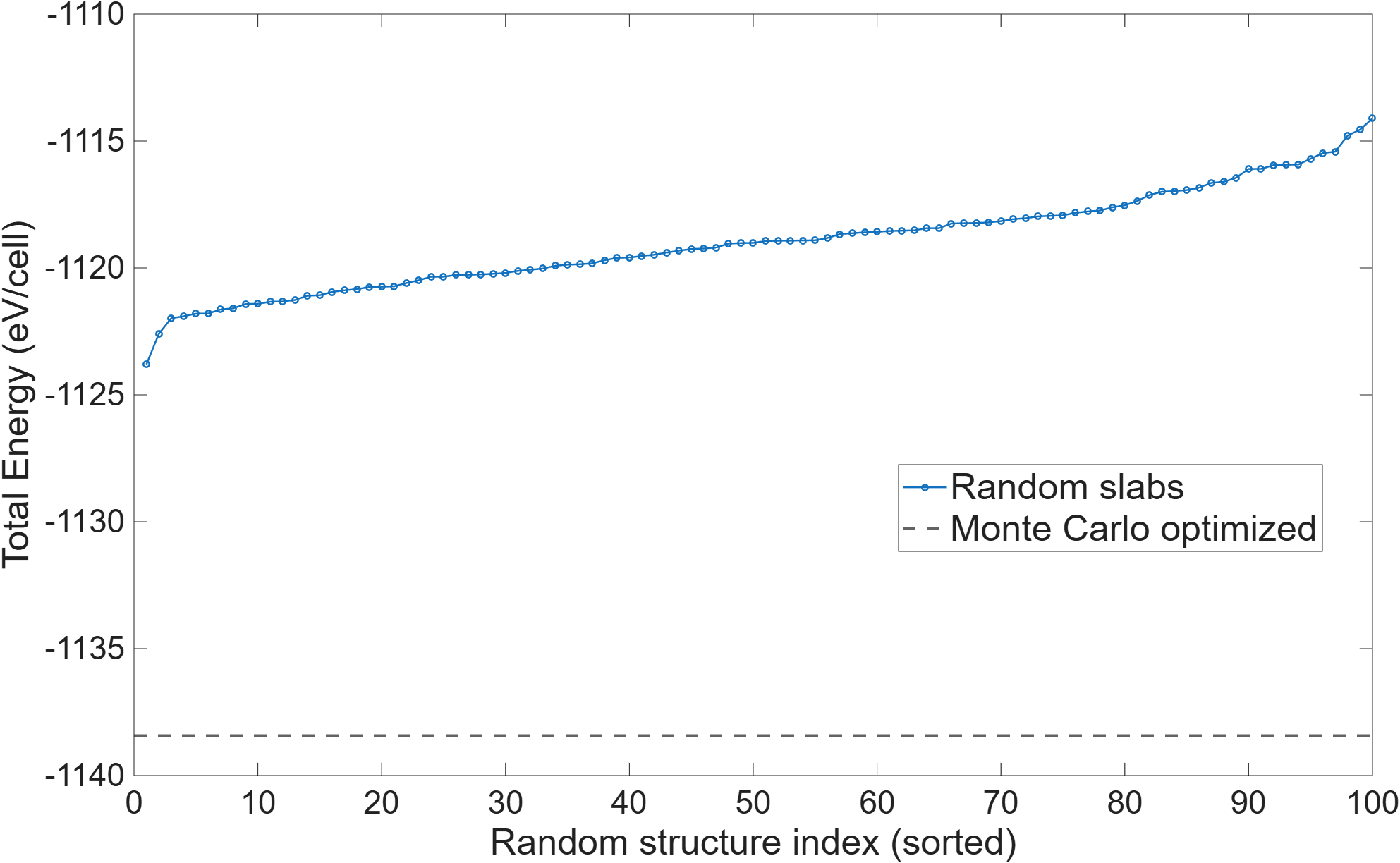}
    \caption{Energies of 100 fully relaxed random configurations of the Ti--V--Cr--Re slab. The Monte Carlo--optimized ground-state energy is marked with a dashed line.}
    \label{fig:slab-random}
\end{figure}

To compare with the conventional random sampling approach used in HEA simulations, we generate 100 random configurations, each fully relaxed and evaluated using the MLIP. Fig.~\ref{fig:slab-random} compares the energies of these configurations (sorted in ascending order) with the Monte Carlo--optimized structure. Even the lowest-energy random configuration remains approximately 15~eV per cell (0.12~eV/atom) higher than the Monte Carlo result, while the average energy gap is about 20~eV per cell. Within the high-dimensional configurational space of this four-component system, the probability of approaching the ground state through uniform random sampling becomes vanishingly small and decreases further with increasing system size. The substantial energy gap confirms that the Monte Carlo--optimized structure reflects a nontrivial ordering pattern inaccessible to random exploration, underscoring the necessity of guided configurational sampling in multi-component alloy systems.
\subsection{Interstitial aggregation and solubility estimation in Nb--Ti--Ta--Hf alloys}
\label{sec:bulk}
Having demonstrated PAIPAI for metal-only configurational sampling, we now extend the framework to systems containing interstitial species and illustrate its use for quantitative thermodynamic analysis based on configurational energetics. Nb--Ti--Ta--Hf-based refractory high-entropy alloys have been widely studied due to their high melting temperatures and favorable mechanical properties, such as strength and room-temperature ductility~\cite{senkov2018effect, liu2022tensile, eleti2019unique}. Previous studies have shown that non-equiatomic Nb-rich variants display excellent formability and fracture resistance~\cite{zhang2023strong}. In addition to metallic configurational disorder, these alloys are highly sensitive to the concentration and distribution of interstitial elements such as oxygen, nitrogen, and carbon, which have been shown experimentally to significantly modify strength and room-temperature ductility~\cite{ko2025boron, jiao2023manipulating, iroc2022design, casillas2020interstitial}.

In this example, we apply PAIPAI to investigate interstitial behavior in the bulk BCC Nb--Ti--Ta--Hf HEA. Starting from a random 250-atom BCC solid-solution structure at a non-equiatomic, Nb-rich composition (Nb$_{113}$Ti$_{63}$Ta$_{37}$Hf$_{37}$), interstitial oxygen or boron atoms are introduced into the octahedral interstitial sites of the BCC lattice. Monte Carlo sampling is performed over both metallic site identities and interstitial occupations, allowing the system to explore low-energy configurations under idealized thermodynamic equilibrium conditions.

\begin{figure}
    \centering
    \includegraphics[width=0.9\linewidth]{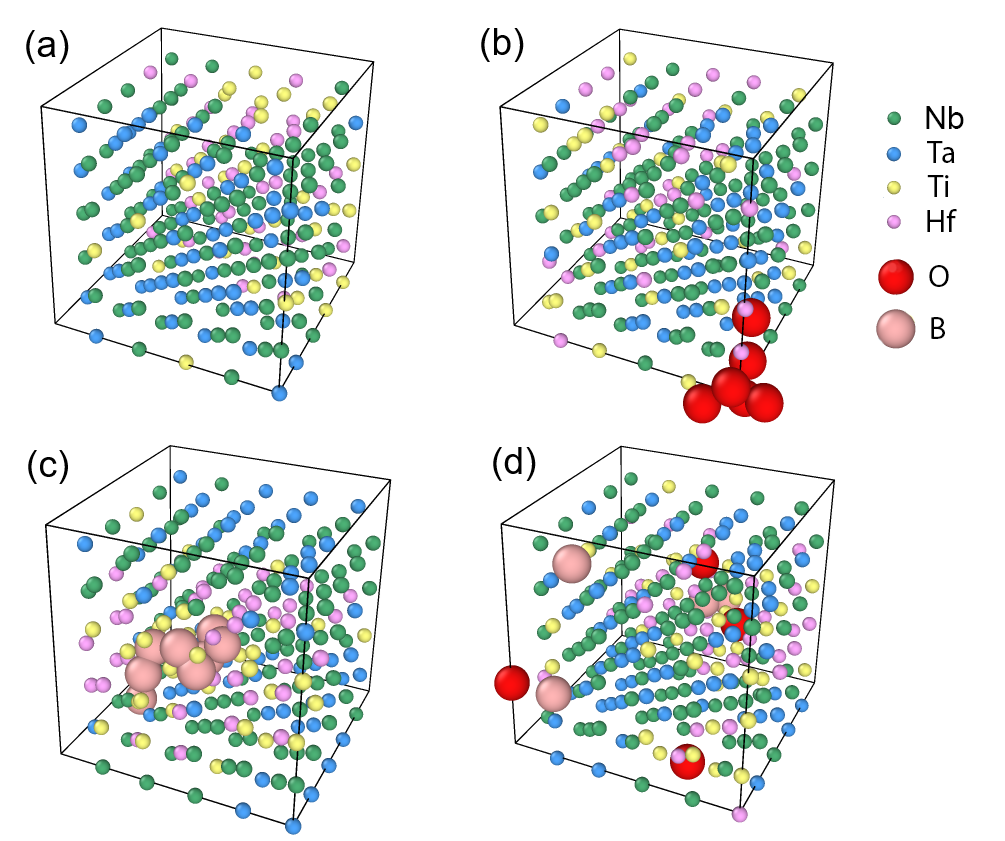}
    \caption{Monte Carlo results of BCC Nb--Ti--Ta--Hf HEA with (a) no interstitials; (b) O interstitials; (c) B interstitials; (d) O and B interstitials. The metallic atoms are plotted with reduced radii to better visualize the configuration.}
    \label{fig:hea-bulk}
\end{figure}

In Fig.~\ref{fig:hea-bulk}(a)--(d), we show the optimized structures of the pure HEA metal, HEA with O interstitials, HEA with B interstitials, and HEA with both B and O interstitials, respectively. In Fig.~\ref{fig:hea-bulk}(a), the pure metallic system exhibits the formation of an Hf-rich BCC region, consistent with the experimentally reported secondary BCC phase enriched in Hf in Nb$_{45}$Ta$_{25}$Ti$_{15}$Hf$_{15}$ alloys~\cite{sahragard2024tensile}. In Fig.~\ref{fig:hea-bulk}(b) and~\ref{fig:hea-bulk}(c), the Monte Carlo sampling reveals that both oxygen and boron tend to aggregate rather than remaining uniformly distributed throughout the lattice. Moreover, they prefer local chemical environments enriched in Hf and Ti, indicating that specific local chemical interactions govern the interstitial energetics. This is consistent with the experimental observation of HfO$_2$ secondary phases in these alloy systems~\cite{huang2026predicting}.
To quantify the thermodynamic driving force associated with interstitial incorporation, we analyze the total energy as a function of the number of interstitial atoms. By evaluating the energy change associated with each additional interstitial, we obtain an effective incremental energy contribution per interstitial at different concentrations, which provides an estimate of the interstitial chemical potential under equilibrium conditions. By comparing the estimated chemical potentials of interstitial species in the BCC matrix with those in relevant competing oxide or boride phases, we can obtain an estimate of the interstitial solubility in the bulk HEA. A detailed example of this analysis for oxygen, including comparison with experimental measurements, was presented in our previous collaborative study~\cite{huang2026predicting}.
%% TODO (Siya): Add reference(s) for experimental agreement. Consider citing the Aomin paper and showing only qualitative boron results here, deferring quantitative boron analysis to the Lavernia collaboration paper.
%% I revised the paragraph.

\subsection{Coupled interstitial and grain boundary segregation in Nb--Ti--Ta--Hf alloys}
\label{sec:gb}
\begin{figure}
    \centering
    \includegraphics[width=0.9\linewidth]{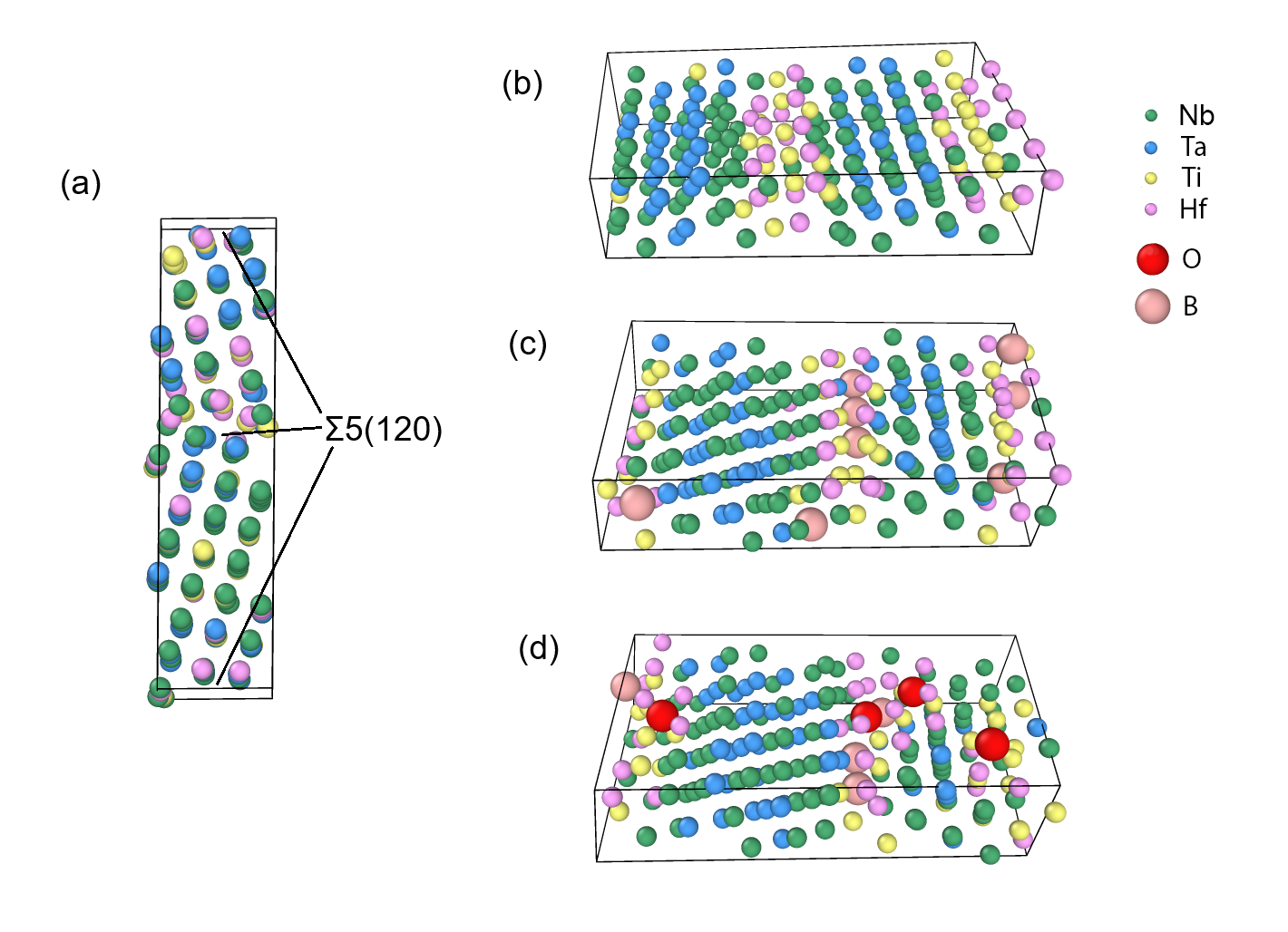}
    \caption{Grain boundary structure of Nb--Ti--Ta--Hf. (a) 200-atom structure with $\Sigma$5(120) grain boundaries in the middle and at both ends; (b) GB structure without interstitials after Monte Carlo; (c) GB structure with B interstitials after Monte Carlo; (d) GB structure with B and O interstitials after Monte Carlo. The metallic atoms are plotted with reduced radii to better visualize the configuration.}
    \label{fig:gb-structure}
\end{figure}

Building upon the previous examples of surface segregation without interstitials and bulk interstitial aggregation without extended defects, we next consider a BCC Nb--Ti--Ta--Hf alloy containing both grain boundaries and interstitial species, thereby simultaneously sampling metallic and interstitial configurational degrees of freedom in the presence of an extended defect. We construct a $\Sigma$5(120) grain boundary structure with 200 metallic atoms in the supercell, with grain boundaries located in the middle and at both ends of the simulation box. Starting from a random metallic configuration, interstitial boron atoms are introduced into the system, and Monte Carlo sampling is performed using PAIPAI. Fig.~\ref{fig:gb-structure} illustrates the grain-boundary structures of (a) the pure alloy; (b) the system containing 8 boron atoms; and (c) the system containing 4 boron atoms and 4 oxygen atoms.
%% TODO (Siya): Consider adding a brief justification for why 200 MC steps are sufficient here (e.g., energy convergence plateau, computational cost), given that the slab example used $10^5$ steps.
%% The slab example searched over 10^5 configurations, but only 105 accepted MC steps. 

For the grain boundary structure with only metallic atoms (Fig.~\ref{fig:gb-structure}(b)), Ti and Hf atoms are observed to segregate to the grain boundary sites, while Nb and Ta remain at bulk-like sites. This is consistent with the segregation of Ti and Hf observed in Fig.~\ref{fig:hea-bulk}(a).

For the grain boundary structure with 8 boron interstitials, the lowest-energy configuration identified by Monte Carlo sampling is illustrated in Fig.~\ref{fig:gb-structure}(c).
In this configuration, Hf and Ti exhibit a pronounced tendency to segregate toward the grain-boundary region, whereas Nb and Ta preferentially occupy bulk-like BCC environments. Boron interstitials are likewise enriched in the vicinity of the grain boundary. From the bulk BCC analysis in Section~\ref{sec:bulk}, we know that B energetically favors neighboring Hf and Ti. This raises a key mechanistic question: is the observed Hf/Ti/B co-segregation primarily driven by intrinsic metallic segregation as observed in Fig.~\ref{fig:gb-structure}(b), or does the presence and grain-boundary segregation of B drive metallic redistribution? In other words, do Hf and Ti intrinsically prefer the grain boundary and subsequently attract B, or does B segregation induce metallic rearrangement?

\begin{figure}
    \centering
    \includegraphics[width=0.9\linewidth]{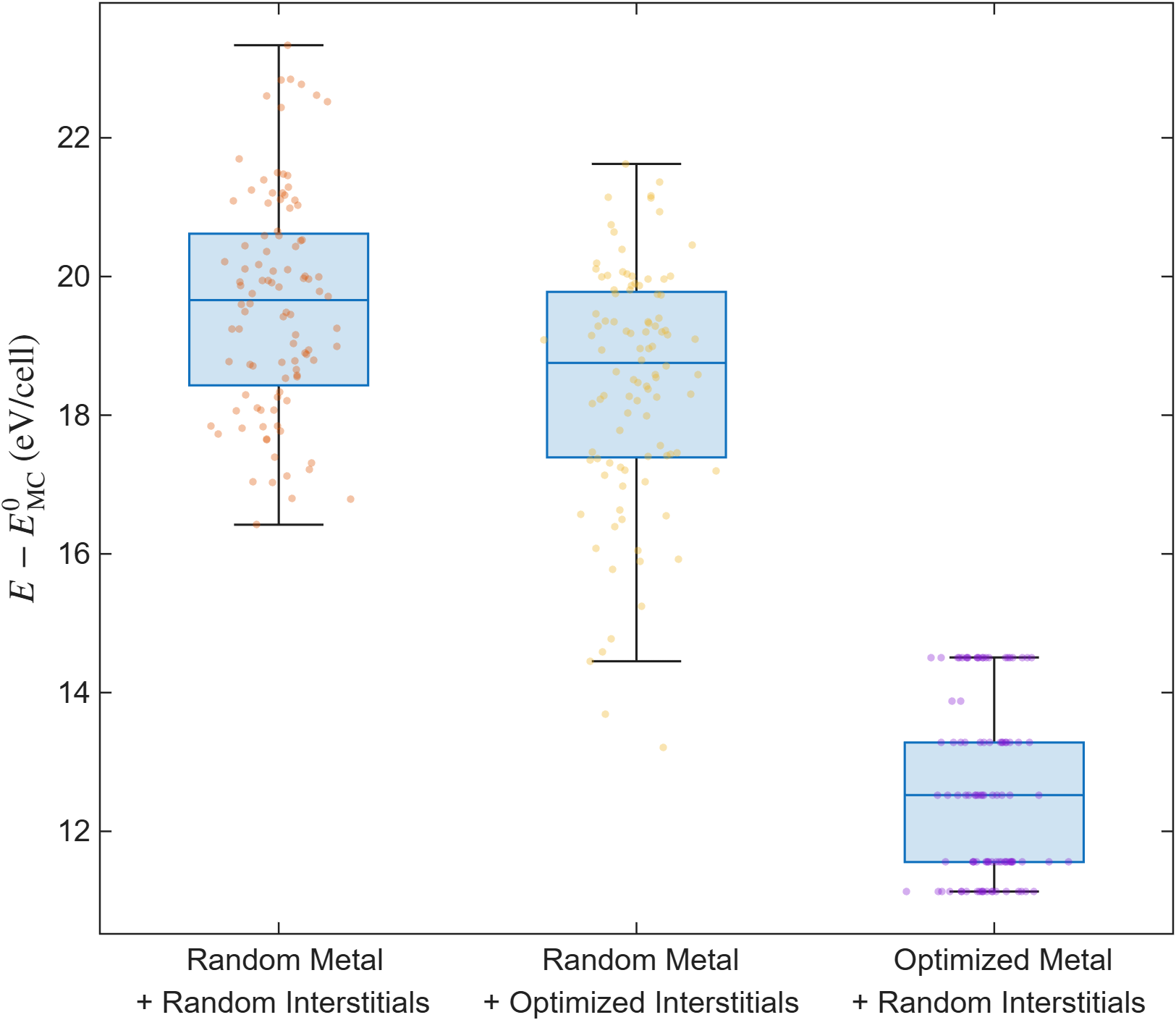}
    \caption{Benchmark of random perturbations applied to the MC-optimized Nb--Ta--Ti--Hf + B grain boundary structure. Three sets of 100 configurations are generated by randomly perturbing (i) both metallic and interstitial atoms; (ii) metallic atoms only; (iii) interstitial atoms only. The box plots show the energy difference relative to the MC-predicted ground state.}
    \label{fig:gb-benchmark}
\end{figure}

To address this question quantitatively, three controlled perturbation tests were performed starting from the Monte Carlo ground state. In the first set, ``random metal + random interstitial,'' the metallic atoms and interstitial atoms were randomly shuffled among lattice sites and interstitial sites, respectively, reproducing conventional random sampling. In the second set, ``random metal + optimized interstitials,'' the optimized B interstitial positions from Fig.~\ref{fig:gb-structure}(c) were fixed while the metallic species were randomly reassigned among lattice sites. In the third set, ``optimized metal + random interstitials,'' the optimized metal configuration from Fig.~\ref{fig:gb-structure}(c) was fixed while the B interstitials were randomly distributed among interstitial sites. For each set, 100 random structures were generated, followed by MLIP relaxation and energy evaluation. The energy differences relative to the MC-optimized structure are plotted in Fig.~\ref{fig:gb-benchmark}.

Fig.~\ref{fig:gb-benchmark} reveals that all three sets of randomly perturbed configurations lie more than 10~eV above the Monte Carlo ground state, reinforcing the conclusion from Section~\ref{sec:slab} that random sampling is insufficient for identifying low-energy configurations. The inadequacy of random sampling is even more pronounced here: the combinatorial spaces associated with the metallic ($2.48\times10^{107}$ possible arrangements) and interstitial ($3.98\times10^{17}$ possible arrangements) degrees of freedom are vastly larger than those of the slab system, making it statistically impossible for 100 random configurations to approach the global minimum. The strongly co-segregated Hf/Ti/B configuration identified by Monte Carlo sampling therefore reflects a highly nontrivial ordering state inaccessible through random exploration.

A more detailed comparison among the three perturbed configuration classes allows a quantitative decomposition of the energetic contributions. The energy difference between ``random metal + random interstitial'' and ``random metal + optimized interstitials'' reflects the energetic contribution of interstitial segregation, while the energy difference between ``random metal + random interstitial'' and ``optimized metal + random interstitials'' reflects the contribution of metallic segregation. As shown in Fig.~\ref{fig:gb-benchmark}, optimizing the interstitial positions alone yields a relatively modest energy reduction, whereas optimizing the metallic distribution produces a substantially larger decrease, confirming that metallic segregation provides the dominant contribution to grain-boundary stabilization.

Importantly, the sum of the individual stabilization effects from metallic and interstitial ordering remains smaller than the total energy difference between the fully random and fully optimized Monte Carlo configurations. This non-additivity indicates that an additional energetic gain arises from the cooperative interaction between Hf/Ti and B segregation at grain boundaries, originating from the energetic preference of B for neighboring Hf and Ti atoms. The results establish a clear causal hierarchy: the intrinsic segregation tendency of Ti and Hf toward the grain boundary plays the primary role in determining the grain-boundary energetics, and the resulting Hf/Ti-enriched boundary regions subsequently provide favorable local coordination environments for B interstitials, thereby facilitating their co-segregation.

In addition, the ``optimized metal + random interstitials'' configurations exhibit significant energetic degeneracy. Once the metallic distribution is fixed at its optimized state, the lattice can be broadly partitioned into two distinct chemical environments: Hf/Ti-enriched grain-boundary regions and Nb/Ta-enriched bulk-like regions. Under such a metallic framework, the energetic contribution of each B interstitial is largely determined by the type of region it occupies, rather than its specific position within that region. As a result, the system energy is governed primarily by the number of B atoms residing in each chemical environment, rather than their detailed spatial arrangement within a given region.

Finally, we consider the case in which both boron and oxygen interstitials are present simultaneously. As shown in Fig.~\ref{fig:gb-structure}(d), both species preferentially segregate to the grain-boundary sites due to the strong attraction of Hf and Ti. This result indicates that the cooperative metal--interstitial segregation mechanism identified above is not limited to a single interstitial species, and that grain boundaries in Nb--Ti--Ta--Hf alloys may act as sinks for multiple light-element impurities simultaneously---a finding with direct implications for understanding embrittlement and secondary-phase nucleation at grain boundaries in refractory HEAs.
%% TODO (Siya): If data are available, consider adding a brief quantitative comparison (e.g., energy reduction relative to the B-only GB system, or relative site occupancy of B vs O at the boundary) to strengthen this paragraph.
%% 

\section{Discussion}
In this work, we introduced PAIPAI, a Monte Carlo--MLIP framework for ground-state structure search in defective HEAs containing interstitial species. By combining Monte Carlo sampling with full structural relaxation, PAIPAI enables direct exploration of the thermodynamically stable atomic ordering in complex HEA systems beyond random solid-solution configurations.

Through three case studies of increasing complexity---surface segregation in a Ti--V--Cr--Re slab, interstitial aggregation in bulk Nb--Ti--Ta--Hf, and coupled metallic--interstitial segregation at a $\Sigma$5(120) grain boundary---we demonstrated the versatility of PAIPAI in treating defective HEA systems containing interstitial species. In each case, the Monte Carlo--optimized configurations were significantly lower in energy than any structure obtained by random sampling, confirming that guided configurational search is essential in the high-dimensional spaces characteristic of multi-component alloys. The use of MLIPs in place of DFT enables full ionic relaxation at every trial step, providing a realistic description of local bonding environments and defect--interstitial interactions at a fraction of the computational cost. Although MLIPs exhibit systematic deviations from DFT energies, we verified that the relative energy ordering of configurations is preserved, ensuring that the identified ground states remain valid at the DFT level.

Although the present study focuses on 0~K ground-state determination, the PAIPAI framework can be extended to finite-temperature configurational thermodynamics. By setting a physical temperature in the Metropolis acceptance criterion and performing equilibrium sampling after convergence while disabling the waiting pool and dual-worker settings, ensemble-averaged properties can be obtained at any target temperature. Configurational entropy is naturally incorporated through canonical sampling, while differences in vibrational free-energy contributions among configurations within the same lattice phase are expected to be relatively small. PAIPAI thus offers a practical framework for assessing both ground-state ordering and temperature-dependent segregation behavior in complex HEAs.

Despite the efficiency of the present multithreaded MC--MLIP framework, convergence can become progressively slower in large systems containing multiple interstitial species and hundreds of atoms. In particular, at later stages of sampling, hundreds of trial configurations may be required to obtain a single accepted move as the system approaches low-energy ordered states. A promising direction to address this bottleneck is adaptive Monte Carlo sampling: rather than generating trial moves with pre-specified probabilities, proposal distributions could be dynamically adjusted based on structural information accumulated during the simulation. For example, if interstitial B atoms are observed to preferentially occupy Hf/Ti-rich octahedral sites, subsequent trial moves could be biased toward similar local motifs. Such adaptive proposal schemes may substantially accelerate convergence in high-dimensional configurational spaces while retaining the rigor of the Monte Carlo framework. More broadly, PAIPAI can be combined with active-learning workflows in which the MLIP is iteratively retrained on configurations encountered during sampling, progressively improving accuracy in the regions of configuration space most relevant to the system of interest.

\section{Methods}
\subsection{Monte Carlo sampling framework}
PAIPAI employs a Monte Carlo framework to explore the configurational space of metallic and interstitial species in defective HEA systems, targeting the identification of low-energy atomic configurations on a fixed lattice topology. At each MC step, a trial configuration is generated by performing either a swap of chemical identities between two metallic sites or a redistribution of interstitial atoms among predefined interstitial sites (e.g., octahedral or tetrahedral). The relative probabilities of metallic and interstitial moves are user-defined, allowing the sampling to be tailored to the system of interest.

The metallic lattice configuration and composition, the coordinates of all possible interstitial sites, and the number of interstitial atoms are specified in the input structure file. For each proposed configuration, structural relaxation and total energy evaluation are performed using the MLIP. The acceptance of a trial configuration follows a Metropolis-type criterion: a trial move with energy change $\Delta E = E_\text{trial} - E_\text{current}$ is accepted with probability
\begin{equation}
P_\text{acc} = \min\!\left(1,\, \exp\!\left(-\frac{\Delta E}{k_B T}\right)\right),
\end{equation}
where $T$ is a user-defined effective temperature parameter and $k_B$ is the Boltzmann constant.

To improve computational efficiency and enable large-scale parallel sampling, PAIPAI adopts a two-tier worker strategy consisting of fast workers and slow workers. Both worker types perform structural relaxation and energy evaluation using the same MLIP model, but differ in their convergence criteria. Fast workers employ loose convergence thresholds ($F_\text{max}=0.1$~eV/\AA\ by default) to rapidly screen candidate configurations at low computational cost, while slow workers apply strict convergence criteria ($F_\text{max}=0.01$~eV/\AA\ by default) to obtain accurate energies suitable for thermodynamic comparison. This separation of roles allows the framework to evaluate a large number of trial configurations without committing full computational resources to each one.

\begin{figure}[htp]
\centering
\includegraphics[width = \columnwidth]{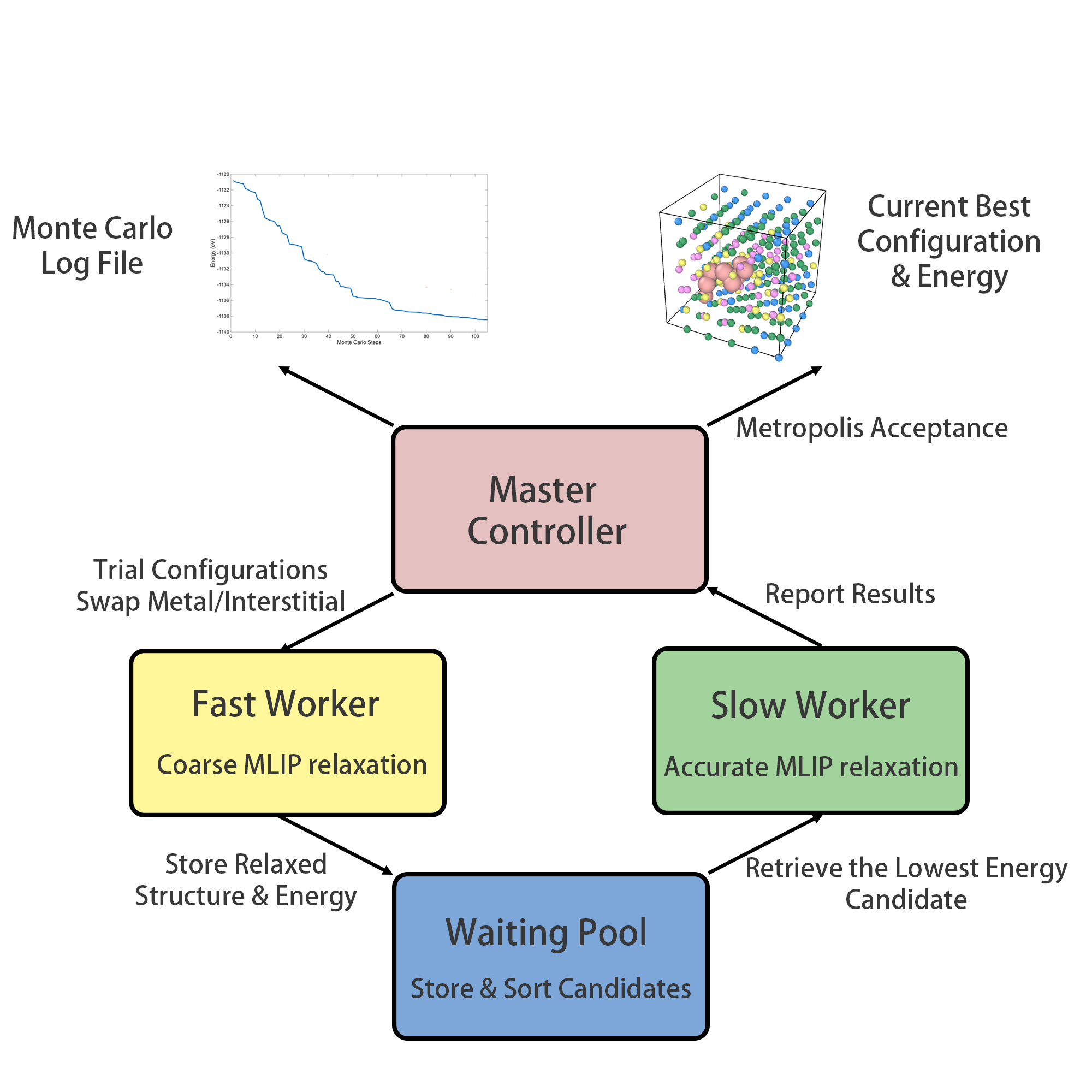}

\caption{Flowchart of the PAIPAI framework illustrating the dual-worker architecture with fast and slow workers coordinated through a shared waiting pool.}
\label{fig:flowchart}
\end{figure}

In the PAIPAI workflow, the master process generates trial configurations through random atomic swaps and dispatches them to available fast workers. The resulting relaxed structures and their corresponding energies are stored in a shared waiting pool. The waiting pool has a fixed capacity (128 by default); when new structures are inserted, the pool is sorted by energy, and only the lowest-energy candidates are retained. Slow workers continuously retrieve the lowest-energy structures from the waiting pool for high-accuracy relaxation. The refined energies are then compared to the current best configuration, and are accepted with a Metropolis acceptance criterion under the specified temperature factor. Once accepted, the current best configuration and energy are updated, and the step is recorded as an accepted MC move. This fast--slow strategy allows more efficient parallel sampling than conventional sequential Monte Carlo approaches. The overall workflow is illustrated in Fig.~\ref{fig:flowchart}. 

In practice, the user should adjust the number of fast and slow workers, as well as their convergence criteria, based on runtime performance, to maximize overall efficiency while maintaining reliability. Specifically, the fast-worker stage should be sufficiently accurate to prevent erroneous rejection of thermodynamically favorable configurations, while maintaining a reasonable population of candidate structures in the waiting pool and avoiding frequent idle time for the slow workers.

In addition, the convergence of the Monte Carlo search is assessed in practice rather than through a simple predefined energetic threshold---the absence of further energy reduction over a limited number of steps does not guarantee that the global minimum has been reached. We recommend prescribing a reasonable wall-time limit and monitoring the evolution of energies along accepted MC steps, together with the structural evolution of the configurations, to assess convergence.

In this work, all simulations are performed on a high-performance computing cluster using 15 fast and 15 slow workers with one CPU core per worker. For pure metallic systems, each system is sampled for 6 days, while systems containing interstitials are sampled for 12 days to ensure adequate convergence.

\subsection{MLIP and DFT calculation settings}
In this work, all MLIP calculations are performed using the GRACE-2L-OMAT model~\cite{bochkarev2024graph}, a graph atomic cluster expansion potential pretrained on the Open Materials (OMat) dataset, which provides broad coverage across the periodic table. The model is implemented within the MaterialsFramework package~\cite{sariturk_2025_15731044}. Although GRACE-2L-OMAT is used throughout this study, the PAIPAI framework is not restricted to a specific potential; alternative MLIP models can be readily substituted by the user.

All DFT calculations are performed using the Vienna \textit{Ab initio} Simulation Package (VASP)~\cite{kresse1996efficient, kresse1993ab, kresse1996efficiency}, with the Perdew--Burke--Ernzerhof (PBE) exchange-correlation functional~\cite{perdew1996generalized} and the projector augmented-wave (PAW) pseudopotentials~\cite{blochl1994projector}. The plane-wave cutoff energy is set to 1.3 times the maximum recommended cutoff energy among the PAW potentials. Brillouin zone integrations are performed using a Monkhorst--Pack mesh with a k-point density of 8000 per reciprocal atom. Spin polarization is included in all calculations. Partial occupancies are treated using first-order Methfessel--Paxton smearing with a smearing width of 0.2 eV. The electronic self-consistency loop is converged to an energy tolerance of $10^{-5}$~eV. Ionic relaxations are carried out using a conjugate-gradient algorithm.

%% TODO (Siya): Add a statement on the convergence criterion for the overall MC simulation — is it a fixed number of steps, an energy plateau, or both? Specify what was used in each case study.

\section*{Data Availability}
All data used in the examples presented in this paper are openly available in the \textit{examples} directory of the repository at \url{https://github.com/siyazhu/PAIPAI}.

\section*{Code Availability}
The \textit{PAIPAI} code is publicly available at \url{https://github.com/siyazhu/PAIPAI}. 

\section*{Competing Interests}
The authors declare no competing interests.

\section*{Acknowledgements}
The authors would like to acknowledge the support of the National Science Foundation through Grant No. 2119103 and 
the U.S. Department of Energy (DOE) ARPA-E ULTIMATE Program through Project DE-AR0001427. RA acknowledges support from the Army Research Laboratory and was accomplished under Cooperative Agreement Number W911NF-22-2-0106. 
Calculations were carried out at the Texas A\&M High-Performance Research Computing (HPRC) Facility.

\section*{Author Contributions}
S.Z. conceived the project, developed the PAIPAI code, performed the calculations, and drafted the manuscript. R.A. provided supervision, project guidance, funding support, and manuscript revision. Both authors reviewed and approved the final manuscript.

\bibliographystyle{elsarticle-num}
\bibliography{ref}

\end{document}